\documentclass[acmsmall,screen,nonacm]{acmart}

\AtBeginDocument{%
	}

\usepackage{svg}
\usepackage{amsmath}
\usepackage{color}
\usepackage{url}
\usepackage{colortbl}
\usepackage{booktabs}
\usepackage{hhline}
\usepackage{multirow}
\usepackage{enumerate}
\usepackage{xspace}
\usepackage{hyperref}
\usepackage{cleveref}
\usepackage{quoting}
\quotingsetup{font={itshape,footnotesize}}
\usepackage{graphicx}
\usepackage{tcolorbox}
\usepackage{subcaption} %  for subfigures environments 
\usepackage[T1]{fontenc}
\usepackage[nolist]{acronym}

\definecolor{lightgray}{rgb}{0.83, 0.83, 0.83}
\definecolor{darkgray}{rgb}{0.66, 0.66, 0.66}
\definecolor{lightergray}{rgb}{0.945,0.945,0.945}

\newcommand{\singlequote}[1]{`#1'}
\newcommand{\doublequote}[1]{``#1''}

\newcommand{\edit}[1]{#1}

\begin{document}
	
	%%
	%% The "title" command has an optional parameter,
	%% allowing the author to define a "short title" to be used in page headers.
	\title{Large Language Models for In-File Vulnerability Localization Can Be “Lost in the End”}
	
	%%
	%% The "author" command and its associated commands are used to define
	%% the authors and their affiliations.
	%% Of note is the shared affiliation of the first two authors, and the
	%% "authornote" and "authornotemark" commands
	%% used to denote shared contribution to the research.
    \author{Francesco Sovrano}
    \orcid{0000-0002-6285-1041}
    \affiliation{%
      \department{Collegium Helveticum}
      \institution{ETH Zurich}
      \city{Zurich}
      \country{Switzerland}
    }
    \affiliation{%
      \institution{University of Zurich}
      \department{Department of Informatics}
      \city{Zurich}
      \country{Switzerland}
    }
    \email{sovrano@collegium.ethz.ch}
    
    \author{Adam Bauer}
    \orcid{0009-0002-5871-1094}
    \affiliation{%
      \institution{University of Zurich}
      \city{Zurich}
      \country{Switzerland}
    }
    \email{adam.bauer@uzh.ch}
    
    \author{Alberto Bacchelli}
    \orcid{0000-0003-0193-6823}
    \affiliation{%
      \institution{University of Zurich}
      \department{Department of Informatics}
      \city{Zurich}
      \country{Switzerland}
    }
    \email{bacchelli@ifi.uzh.ch}
	
	%%
	%% By default, the full list of authors will be used in the page
	%% headers. Often, this list is too long, and will overlap
	%% other information printed in the page headers. This command allows
	%% the author to define a more concise list
	%% of authors' names for this purpose.
	\renewcommand{\shortauthors}{Sovrano et al.}
	
	%%
	%% The abstract is a short summary of the work to be presented in the
	%% article.
	\begin{abstract}
		Traditionally, software vulnerability detection research has focused on individual small functions due to earlier language processing technologies' limitations in handling larger inputs. However, this function-level approach may miss bugs that span multiple functions and code blocks. Recent advancements in artificial intelligence have enabled processing of larger inputs, leading everyday software developers to increasingly rely on chat-based large language models (LLMs) like GPT-3.5 and GPT-4 to detect vulnerabilities across entire files, not just within functions. 
		This new development practice requires researchers to urgently investigate whether commonly used LLMs can effectively analyze large file-sized inputs, in order to provide timely insights for software developers and engineers about the pros and cons of this emerging technological trend. Hence, the goal of this paper is to evaluate the effectiveness of several state-of-the-art chat-based LLMs, including the GPT models, in detecting in-file vulnerabilities.
		We conducted a costly investigation into how the performance of LLMs varies based on \emph{vulnerability type}, \emph{input size}, and \emph{vulnerability location} within the file.
		To give enough statistical power ($\beta \geq .8$) to our study, we could only focus on the three most common (as well as dangerous) vulnerabilities: XSS, SQL injection, and path traversal.
		Our findings indicate that the effectiveness of LLMs in detecting these vulnerabilities is strongly influenced by both the location of the vulnerability and the overall size of the input. Specifically, regardless of the vulnerability type, LLMs tend to significantly ($p < .05$) underperform when detecting vulnerabilities located toward the end of larger files---a pattern we call the \singlequote{lost-in-the-end} effect. 
		Finally, to further support software developers and practitioners, we also explored the optimal input size for these LLMs and presented a simple strategy for identifying it, which can be applied to other models and vulnerability types. Eventually, we show how adjusting the input size can lead to significant improvements in LLM-based vulnerability detection, with an average recall increase of over 37\% across all models. % reaching up to 95\%.
        \\
        \textbf{Replication Package:} \href{doi.org/10.5281/zenodo.14840519}{https://doi.org/10.5281/zenodo.14840519}
	\end{abstract}
	
	%%
	%% The code below is generated by the tool at http://dl.acm.org/ccs.cfm.
	%% Please copy and paste the code instead of the example below.
	%%
	
	
	%%
	%% Keywords. The author(s) should pick words that accurately describe
	%% the work being presented. Separate the keywords with commas.
	%% A "teaser" image appears between the author and affiliation
	%% information and the body of the document, and typically spans the
	%% page.
	%\begin{teaserfigure}
	%  \includegraphics[width=\textwidth]{sampleteaser}
	%  \caption{Seattle Mariners at Spring Training, 2010.}
	%  \Description{Enjoying the baseball game from the third-base
		%  seats. Ichiro Suzuki preparing to bat.}
	%  \label{fig:teaser}
	%\end{teaserfigure}
	
	%%
	%% This command processes the author and affiliation and title
	%% information and builds the first part of the formatted document.
	\maketitle
	
	\begin{acronym}
		\acro{AI}{artificial intelligence}
		\acro{LLM}{large language model}
		\acro{RAG}{retrieval-augmented generation}
		\acro{IDE}{integrated development environment}
		\acro{CWE}{Common Weakness Enumeration}
		\acro{CVE}{Common Vulnerabilities and Exposures}
		\acro{XSS}{cross-site scripting}
	\end{acronym}
	
	\section{Introduction} \label{sec:introduction}
	
	%%%%%%%%%%%%%%%%%%
	%%% The Context and Motivation
	%%%%%%%%%%%%%%%%%%
	In the rapidly evolving field of software development, the integration of \acp{LLM} has shifted from a novel concept to a mainstream practice. As generative \ac{AI} technologies continue to advance, tools like ChatGPT, Copilot, and similar models are now embedded as first-class citizens into development environments, assisting with tasks such as code generation, bug fixing, and security analysis. According to Sonatype’s 2023 report \cite{sonatype2023state}, 97\% of developers and security professionals now rely on \acp{LLM} in their workflows, marking a significant shift in how software vulnerabilities are detected and patched.
	
	Indeed, the integration of \acp{LLM} addresses a critical challenge identified in the 2022 GitLab Survey \cite{gitlab2022devsecops}, which noted that developers often fail to detect security-relevant bugs early and do not prioritize bug fixing adequately. 
	This typically leads to frequent vulnerabilities that are challenging for humans to detect. 
	Instead, \acp{LLM} have been found to be quite effective in supporting vulnerability detection \cite{ullah2024llms,DBLP:conf/apsec/FuTNL23,DBLP:journals/corr/abs-2309-05520,DBLP:journals/corr/abs-2401-15468}. This efficacy is primarily attributed to their capability to interpret code semantics \doublequote{like a human} \cite{nam2024using}, an advantage not typically possessed by fuzzers and classical static analyzers.
	
	Historically, the detection of software vulnerabilities has primarily focused on analyzing individual functions due to limitations in earlier natural language processing and machine learning models \cite{DBLP:journals/corr/abs-2401-15468,DBLP:conf/kbse/ZhouKXLHL23,DBLP:journals/ese/NapierBW23,DBLP:conf/raid/0001DACW23,DBLP:journals/ese/NapierBW23,cheshkov2023evaluation,DBLP:journals/corr/abs-2308-12697,LIU2024107458}. However, this narrow scope can overlook vulnerabilities that span across multiple functions or different parts of a file (as in the example shown in Figure \ref{fig:cve_fix}), leading to an incomplete or missed detection \cite{vassallo2018context}.
	
	Recent advancements in \acp{LLM} have significantly expanded their capacity to process large portions of code, now reaching lengths of millions of characters \cite{DBLP:journals/corr/abs-2308-10882,DBLP:journals/corr/abs-2402-13753,DBLP:journals/corr/abs-2306-15595,DBLP:journals/corr/abs-2401-03462}. These models can now analyze entire files or even entire repositories. As a result, software developers are increasingly relying on chat-based \acp{LLM} like GPT-3.5 and GPT-4 to detect vulnerabilities across whole files. Given this growing reliance, it is crucial to assess whether these commonly used state-of-the-art chat-based \acp{LLM} can effectively identify vulnerabilities in large, file-sized inputs. Such an evaluation will provide software developers and engineers with timely insights into the strengths and limitations of this emerging technology, helping them make informed decisions about its use.

    \edit{In-file vulnerability detection is crucial because many vulnerabilities involve multiple functions. Of the top-10 most dangerous security vulnerabilities in 2024 \cite{CWE2024Top25}, at least six frequently span functions: \singlequote{Out-of-bounds Write} (\#2) and \singlequote{Out-of-bounds Read} (\#6) involve buffer allocation and bounds checking in different file areas, \singlequote{Cross-site Scripting}, \singlequote{SQL Injection}, and \singlequote{Path Traversal} (\#1, \#3, \#5) span input sanitization handling across modules, \singlequote{Use After Free} (\#8) involves memory allocation, freeing, and reuse in separate areas, and \singlequote{Missing Authorization} (\#9) spans missing authorization checks across access points.}
	
	This leads to important questions: Can \acp{LLM} safely detect vulnerabilities within files that are up to the size of their context windows, or do intrinsic limitations exist that necessitate the use of smaller inputs? If smaller inputs are required, how small must they be? This study seeks to answer these questions, offering practitioners a clearer understanding of the capabilities and limitations of \acp{LLM} in detecting vulnerabilities within large code-bases.
	
	%%%%%%%%%%%%%%%%%%
	%%% The Study
	%%%%%%%%%%%%%%%%%%	
	We thus conducted several empirical studies involving open-source \acp{LLM} like Mixtral 8x7b, Mixtral 8x22b, and Llama 3 70B, as well as well-known commercial models such as GPT-3.5, GPT-4, and GPT-4o. These chat-based models were employed off-the-shelf (i.e., without customization, as a typical software developer might use them) and tested across a range of scenarios involving real-world security vulnerabilities from the public \ac{CVE} catalog \cite{CVE}, which lists software vulnerabilities categorized under the \ac{CWE}, a standardized system for identifying software weaknesses. %, which comprises over 230,000 records. 
	
	To assess the vulnerability detection capabilities of these \acp{LLM} in this realistic setting, we used a balanced dataset, which we compiled from the \ac{CVE} catalog by following established best practices in the vulnerability localization literature \cite{DBLP:journals/ese/NapierBW23,DBLP:conf/raid/0001DACW23,cheshkov2023evaluation,DBLP:journals/corr/abs-2308-12697,ullah2024llms,wang2024reposvul}. Specifically, the dataset we assembled comprises over 750 files containing vulnerabilities, along with over 750 \doublequote{non-vulnerable} files obtained by applying the patches for these vulnerabilities as provided in the \ac{CVE} catalog.
	
	We ensured that the dataset was constructed such that each file represented a unique vulnerability instance. Before expanding the study to multi-file systems or entire repositories, we first focused on single-file vulnerability detection.
	
	Additionally, to achieve sufficient statistical power ($\beta \geq .8$) in our study, we considered only three common CWE types: \ac{XSS}, SQL injection, and path traversal. These were the only CWE types for which we identified at least 100 distinct vulnerability instances, a number determined by \textit{a priori} power analysis \cite{erdfelder1996gpower}.
	
	We then used this dataset to measure the effectiveness of each \ac{LLM} at detecting vulnerabilities using precision, recall, accuracy, and F1 score metrics.
	Afterwards, we examined how file size and the position of the vulnerability within the file impact the effectiveness of the \acp{LLM}. This part of the study involved two key experiments. The first experiment assessed the correlation between input size and the models' detection accuracy across the over 750 real-world vulnerabilities we collected from the \ac{CVE} catalog. The second experiment, referred to as the \singlequote{code-in-the-haystack} experiment, specifically investigated how relocating the same vulnerable lines of code within files of varying sizes affects the models' detection capabilities. Through these experiments, we identified and quantified a \singlequote{lost-in-the-end} issue, where well-known \acp{LLM} such as ChatGPT tend to miss vulnerabilities located at the end of long files.
	
	This finding contrasts with the common trend identified in the literature, where LLMs are often observed to suffer from a \singlequote{lost-in-the-middle} issue (i.e., they process information well at the beginning and end, but not in the middle), rather than a \singlequote{lost-in-the-end} problem \cite{li2024snapkv,DBLP:journals/corr/abs-2404-08865,liu2024lost,an2024make,xu2023retrieval}.
	While we did not analyze additional types of vulnerabilities, the trends observed suggest that the \singlequote{lost-in-the-end} issue may also affect other vulnerability types. Regardless, the three vulnerabilities we studied are among the most prevalent and dangerous in web security, making our findings highly relevant to the community.
	
	Building on these findings, we empirically investigated how small an input code needs to be to maximize the probability of detecting a vulnerability with one of the considered \acp{LLM}, thereby mitigating the \singlequote{lost-in-the-end} issue. This involved using simple file chunking strategies to adjust the dataset file sizes and analyzing how these changes influenced the models' ability to correctly identify vulnerabilities. By observing the resulting detection scores, we identified more effective input configurations for each \ac{LLM} and vulnerability type. 
	
	This knowledge should help practitioners better understand how to effectively use \acp{LLM}, such as ChatGPT and Llama, in practice for vulnerability detection.
	
	% %%%%%%%%%%%%%%%%%%
	% %%% The Contributions
	% %%%%%%%%%%%%%%%%%%
	%    Finally, our main contribution is a comparative analysis of the effectiveness of commercial and open-source state-of-the-art \acp{LLM} in locating vulnerabilities within files, also considering characteristics like \ac{CWE} type and frequency.
	% % Our main contributions include:
	% % \begin{itemize}
		% % 	\item A comparative analysis of the effectiveness of commercial and open-source state-of-the-art \acp{LLM} in locating vulnerabilities within files, also considering characteristics like \ac{CWE} type and frequency.
		% % 	\item A novel code-in-the-haystack method to study the lost-in-the-end issue of \acp{LLM} on code.
		% % 	\item An investigation and strategy on how to increase the recall of off-the-shelf \acp{LLM} in vulnerability detection tasks.
		% % 	\item A dataset~\cite{anonymous2024infile} of vulnerabilities in PHP, TypeScript, JavaScript, HTML, Java, Go, Python, Ruby, and C, extracted from the \ac{CVE} catalog, which can be used to assess \acp{LLM}' effectiveness in detecting vulnerabilities within files.
		% % \end{itemize}
	
	\section{Background and Related Work} \label{sec:related_work} \label{sec:background}
	
	This section compares our study with other research on language models for vulnerability detection and explores the relationship to prior natural language processing research, focusing on the \singlequote{lost-in-the-middle} issue and how input size affects \acp{LLM} performance.
	
	\subsection{Vulnerability Detection via LLMs} \label{sec:related_work:bug_detection} \label{sec:background:cwe_n_cve}
	
	There has been extensive research on the application of language models for vulnerability detection in software development and maintenance, primarily involving small specialized models operating at a function-level granularity \cite{DBLP:journals/corr/abs-2401-15468,DBLP:conf/kbse/ZhouKXLHL23,DBLP:journals/ese/NapierBW23,DBLP:conf/raid/0001DACW23,cheshkov2023evaluation,DBLP:journals/corr/abs-2308-12697,LIU2024107458}. The recent advent of ChatGPT has shifted some focus to the use of \acp{LLM}, but still at a function-level granularity \cite{DBLP:journals/corr/abs-2401-15468,steenhoek2024comprehensive,purba2023software,DBLP:conf/acsac/ThapaJACPN22}.
	
\begin{figure}[tbh]
		\centering
		\includegraphics[width=.95\columnwidth]{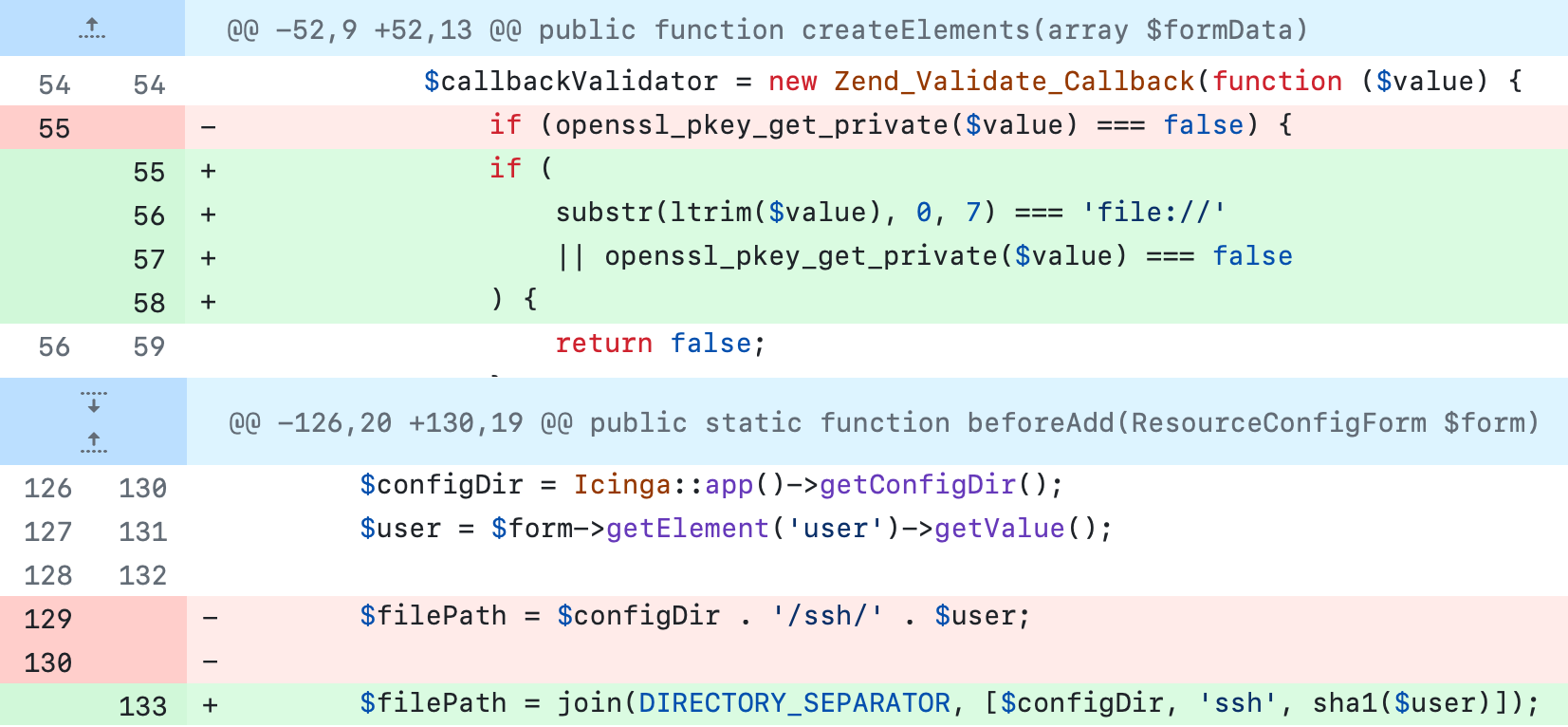}
		\caption{GitHub commit for CVE-2022-24715, showing a security patch to prevent a path traversal.}
		\label{fig:cve_fix}
	\end{figure}

	The majority of the aforementioned studies used vulnerability datasets extracted from the \ac{CVE} catalog, a practice we also adopt in our research.
	The \ac{CVE} MITRE catalog is a public database managed by the MITRE Corporation that lists and describes software and firmware vulnerabilities, assigning them unique identifiers for standardized handling. The catalog classifies vulnerabilities using the \acl{CWE}, which categorizes software and hardware weaknesses with security implications. Each \ac{CWE} entry details a specific type of vulnerability, helping organizations to better identify, understand, and mitigate potential threats. An example of \ac{CVE} entry of type CWE-22, wherein the fix spans multiple functions. is shown in Figure \ref{fig:cve_fix}.
	
	Studies such as the one by \citet{ullah2024llms} have shown significant non-robustness in advanced models like PaLM2 and GPT-4. Changes in function or variable names, or the addition of library functions, can cause these models to yield incorrect answers in 26\% and 17\% of cases, respectively. This highlights the need for further advancements before \acp{LLM} can serve as reliable security assistants. Although \citeauthor{ullah2024llms}'s work \cite{ullah2024llms} studies code augmentation, unlike ours it does not explore the relationship between bug positions and the \acp{LLM}'s ability to identify vulnerabilities.
	
	Instead, \citet{DBLP:journals/corr/abs-2309-05520} found that ChatGPT's performance varies across different vulnerabilities and bugs, and multi-round conversations often worsens detection rates. However, they do not investigate the underlying reasons for this, which we attribute to the \singlequote{lost-in-the-end} problem, where more context leads to information being lost.
	
	%In contrast, \cite{cheshkov2023evaluation} report poor performance of ChatGPT in vulnerability localization. Our results show better performance, potentially due to the use of a more recent version of ChatGPT or a more effective prompting strategy. It is noteworthy that \cite{cheshkov2023evaluation} is an arXiv preprint and has not been peer-reviewed.
	
	A more recent work by \citet{DBLP:journals/corr/abs-2401-15468} has shown that with specific prompts (containing examples of the bug), ChatGPT can outperform smaller specialized models like CodeBERT \cite{DBLP:conf/emnlp/FengGTDFGS0LJZ20} and CodeT5 \cite{DBLP:conf/emnlp/0034WJH21} in detecting vulnerabilities. However, their analysis is limited to function-level detection and simple yes-or-no questions without explicit localization of the main line of code responsible for the vulnerability. Our study extends to in-file detection, using post-fix code as negatives, making the detection task more challenging and realistic.
	
	%\cite{DBLP:journals/corr/abs-2401-15468} has shown that \acp{LLM} like ChatGPT 4 can locate top-25 \ac{CWE} vulnerabilities quite well, however their experiments were conducted in a completely different way than ours. Indeed, they performed vulnerability localization only at a function level, moreover, their prompt was only requiring a yes-or-no answer with no explicit localization of the main line of code responsible for the vulnerability. Moreover, the way they generated the negatives (i.e., the data points without any vulnerability) was completely different and, we believe, prone to increasing the true negatives rate due to the fact that they may have considered functions that do not contain any possible reference to code related to a specific bug. For example, feeding to a \ac{LLM} functions that do not contain any SQL query and asking to the model if there may be any SQL injection (CWE-89) is very likely to push the \ac{LLM} to say no. Instead, as the negatives we use the files right after the fix patch, this means that we feed code which is related to a given CWE and therefore harder to spot as not a vulnerability.
	
	\citet{DBLP:journals/corr/abs-2401-15468} try several prompts to understand their impact in vulnerability detection, also suggesting that vulnerabilities belonging to less common \ac{CWE} types (e.g., CWE-22; path traversal) are also less likely to be identified by \acp{LLM}.
	Indeed, according to \citet{DBLP:conf/kbse/ZhouKXLHL23}, vulnerability data exhibit a long-tailed distribution in terms of \ac{CWE} types: a small number of \ac{CWE} types have a substantial number of samples (e.g., CWE-79; \ac{XSS}), while numerous \ac{CWE} types have very few samples. However, with our experiments we empirically show that with \acp{LLM}, the most frequent weaknesses (e.g., CWE-79) are actually those less likely to be identified by the model compared to less frequent ones (e.g., CWE-22).
	
	\subsection{\singlequote{Lost-in-the-Middle} Issue and Input Size Impact on LLMs} \label{sec:related_work:lost_in_the_middle} \label{sec:background:needle_in_the_haystack}
	
	%One important reason for the code context not being correctly captured by a \ac{LLM} is that, as shown by the experiments of \citet{DBLP:journals/corr/abs-2402-14848}, the ability of \acp{LLM} to reason over an input text tends to be inversely proportional to the text size. 
	Although many modern \acp{LLM}, including ChatGPT, can process extensive inputs, they often struggle to effectively manage information within lengthy contexts. 
	This challenge has been highlighted in many other studies sometimes referring to a \singlequote{lost-in-the-middle} issue \cite{liu2024lost}. 
	\singlequote{Lost-in-the-middle} refers to a phenomenon in large language models where information presented in the middle of a long input sequence is more likely to be overlooked or forgotten compared to information at the beginning or end.
	
	These problems seem to happen because these language models do not know well how to identify relevant details amid fairly large contexts, and this has an impact on their ability to consistently apply contextual knowledge over extended sequences of information, reasoning over code and locating vulnerabilities.
	
	To identify and analyze the \singlequote{lost-in-the-middle} issue, researchers have recently started to use the \singlequote{needle-in-the-haystack} experiment.
	In this approach, a specific piece of information, known as the \singlequote{needle}, is hidden within a large block of irrelevant or filler text, referred to as the \singlequote{haystack}. The model is then tasked with finding and retrieving this specific information.
	This method was recently popularized online by specialized blogs \cite{author2023needle} and it is now used to assess the memory capabilities of \acp{LLM}.
	Since its introduction, several gray-literature papers (i.e., non-peer-reviewed arXiv preprints) have adopted the needle-in-the-haystack method to evaluate LLMs' ability to recall information and effectively use input memory. Notable examples include works by \citet{li2024snapkv,DBLP:journals/corr/abs-2404-08865,liu2024lost,an2024make,xu2023retrieval}.
	
	Typically, the standard needle-in-the-haystack experiment only requires a \ac{LLM} to identify a specific piece of information within the input and does not involve reasoning over the entire content, as required in vulnerability detection.
	Therefore, we performed an adaptation (which we named \singlequote{code-in-the-haystack}) of this approach specifically for vulnerability localization, requiring reasoning over an entire code context. This context is not just random text or code, but related code taken from the same file and repository.
	
	Our results indicate that the \acp{LLM} we considered for vulnerability localization do not really suffer from the \singlequote{lost-in-the-middle} issue but rather from a \singlequote{lost-in-the-end} issue.
	This is aligned with the findings of \citet{DBLP:journals/corr/abs-2402-14848} which show that large input sizes can hinder the reasoning capabilities of \acp{LLM}.
	However, \citet{DBLP:journals/corr/abs-2402-14848} focused on general reasoning tasks unrelated to coding, whereas our study builds on this by exploring the impact of input size on the ability of \acp{LLM} to detect specific kinds of vulnerabilities within large code contexts.
	
	\section{Research Questions} \label{sec:study_design} \label{sec:research_questions}
	
	Recent advancements in artificial intelligence have made it possible for \acp{LLM} to process entire files or even repositories, rather than just small code segments. As the popularity of chat-based \acp{LLM} continues to grow and the use of these \acp{LLM} for in-file vulnerability detection becomes more widespread among software developers, it is crucial to assess their effectiveness across various vulnerability types to inform practitioners of the pros and cons of these technological trends. 
	This urgency led us to our initial research question:
	
	\begin{tcolorbox}[boxsep=1mm, top=1mm, bottom=1mm] % [boxsep=1mm, top=1mm, bottom=1mm, after=\vspace{-0.5mm}]
		\textbf{RQ1: To what extent can popular chat-based \acp{LLM} detect vulnerabilities within entire files?}
	\end{tcolorbox}
	
	To address this question, we first obtained a dataset containing an equal distribution of vulnerable and patched (i.e., without the vulnerability) files for different \ac{CWE} types.
	As detailed in Section \ref{sec:data}, the dataset was extracted from the \ac{CVE} catalog (which includes known vulnerabilities and their fixes/patches; cf. Section \ref{sec:background:cwe_n_cve}), by selecting all vulnerabilities wherein the patch targets a single file of code.
	Following the methodology described in Section \ref{sec:rq1}, we then instructed popular state-of-the-art \acp{LLM} such as GPT-4 to find and return any line within the dataset files that may lead to a vulnerability of a given \ac{CWE} type. This approach allowed us to compute metrics such as precision, recall, accuracy, and F1 score for each \ac{LLM} and \ac{CWE} type.
	
	%    This procedure of extracting the dataset from the \ac{CVE} catalog using known vulnerabilities and their corresponding patches as the \doublequote{non-vulnerable} version is well-established in the vulnerability localization literature \cite{DBLP:journals/ese/NapierBW23,DBLP:conf/raid/0001DACW23,cheshkov2023evaluation,DBLP:journals/corr/abs-2308-12697,ullah2024llms,wang2024reposvul}, despite the impossibility of mathematically guaranteeing that the post-fix code is fully bug-free. This impossibility stems from Rice's theorem \cite{rice1953classes}, which demonstrates that, in the most general case, it is undecidable whether a program is free of vulnerabilities. Nevertheless, we are certain that the post-fix code resolves at least one vulnerability, and that this vulnerability must be detected by the LLM.
	
	Once we determine the effectiveness of these models in detecting vulnerabilities within files, we can investigate the reasons behind their errors to inform practitioners about best practices. To do this, we decided to examine whether the position of the vulnerability or the file size impacts the LLMs' performance, specifically assessing whether they experience any \singlequote{lost-in-the-middle} issues (cf. Section \ref{sec:background:needle_in_the_haystack}). This leads to our next question:
	
	\begin{tcolorbox}[boxsep=1mm, top=1mm, bottom=1mm]
		\textbf{RQ2: To what extent input size and vulnerability position affect \acp{LLM}' ability to detect vulnerabilities?}
	\end{tcolorbox}
	
	To answer it, we conducted two experiments. 
	The first experiment (cf. Section \ref{sec:rq2:exp1}) used the dataset from RQ1 to examine the correlation between input size, the position of the vulnerability, and the correctness of the \ac{LLM}'s output, focusing only on the files containing a vulnerability. 
	
	However, this first experiment presented a problem: the distribution of actual file sizes and vulnerability locations follows a tailed distribution, and this could distort the results.
	Therefore, we conducted a second experiment (whose methodology is detailed in Section \ref{sec:rq2:exp2}), which complements the previous one by uniforming the distribution of vulnerability locations and file sizes. Consequently, this second experiment was designed to more effectively determine whether a given \ac{LLM} suffers from a \singlequote{lost-in-the-middle} or \singlequote{lost-in-the-end} problem.
	
	Finally, building on the insights from RQ2, we aimed to establish best practices for practitioners when using mainstream LLMs in cases where they suffer from the \singlequote{lost-in-the-middle} issue or similar problems. If these LLMs have intrinsic limitations that hinder the safe detection of vulnerabilities in files approaching the size of their context windows, we seek to provide practitioners with guidance on the maximum input size that still ensures reliable detection capabilities. This led to the formulation of our third research question:
	
	\begin{tcolorbox}[boxsep=1mm, top=1mm, bottom=1mm]
		\textbf{RQ3: If smaller inputs are required, how small must they be?}
	\end{tcolorbox}
	
	We hypothesize that the optimal input size could be determined by analyzing how variations in input size affect the \ac{LLM}'s ability to accurately identify the location of a vulnerability. To test this, we divided the vulnerable files into chunks of varying sizes, following the methodology presented in Section \ref{sec:rq3} and adapting the experimental method from RQ1 to assess file chunks rather than entire files. After determining the optimal input size for an \ac{LLM} and \ac{CWE} type, practitioners can chunk files accordingly, improving vulnerability detection likelihood within their code.
	
	%By answering these research questions, we aim to advance the understanding of how \acp{LLM} perform in file-level vulnerability detection and identify factors that can enhance their accuracy and reliability.
	
	\section{Dataset Construction} \label{sec:rq1:data} \label{sec:data}
	
	To address RQ1, we need a dataset with an equal distribution of vulnerable and non-vulnerable files, categorized by CWE type, where each vulnerability fix affects only a single file. This ensures that the LLM has sufficient context to understand the vulnerability. As no such dataset was available in the related literature (see Section \ref{sec:related_work}), we created one ourselves by extracting data from the CVE catalog, as has been done in related studies \cite{DBLP:journals/ese/NapierBW23,DBLP:conf/raid/0001DACW23,cheshkov2023evaluation,DBLP:journals/corr/abs-2308-12697,ullah2024llms,wang2024reposvul}. Indeed, the CVE catalog contains thousands of entries, each linking to a specific vulnerability and its corresponding patch \edit{(i.e., the vulnerability fix)}.
	
	\textbf{Methodology.} Following an established procedure in vulnerability localization research \cite{DBLP:journals/ese/NapierBW23,DBLP:conf/raid/0001DACW23,cheshkov2023evaluation,DBLP:journals/corr/abs-2308-12697,ullah2024llms,wang2024reposvul}, we labeled the file before the patch as the \doublequote{vulnerable} file and the file after the patch as the \doublequote{non-vulnerable} file. This ensures that the pre-fix code contains a vulnerability that the LLM must detect, while the post-fix code should be free of the same vulnerability. Although Rice’s theorem \cite{rice1953classes} suggests it is impossible to determine whether a piece of code is entirely free of vulnerabilities, we use these \doublequote{non-vulnerable} files only for answering RQ1, which benchmarks LLMs. This does not affect our core findings and contributions which stem from RQ2 and RQ3.
	
	We initiated the data extraction process by scraping all \ac{CVE} entries from March 2018 to March 2024, covering a period of six years. 
	These entries were filtered to include only those referring to single GitHub commits affecting a single source code file, excluding documentation files. For instance, if a commit targeted one code file and two documentation files, it was included, whereas commits containing multiple code files were not.
	Eventually, we only considered files written in: PHP, TypeScript, JavaScript, HTML, Java, Go, Python, Ruby, and C.
	
	Consequently, we excluded file extensions associated with documentation or data, such as txt, svg, md, xml, and json. We also excluded file extensions associated with C++, including header files.
	Since RQ1 is on in-file vulnerability localization, this focus on single-file changes ensures that the quality of the findings is not compromised by the fact we did not consider more files per vulnerability, such as entire repositories or C++ software (comprising both .cpp and .h files).
	
	Since this dataset will be used for several statistical tests (e.g., logistic regressions), we included only \ac{CWE} types with at least 100 \ac{CVE} entries to ensure the tests have sufficient power ($\beta \geq .8$), as indicated by an \textit{a priori} power analysis \cite{erdfelder1996gpower}. This analysis assumes a uniform distribution of bug positions, a large effect size of bug position or file size on vulnerability detection, and an $\alpha < .05$.
	This choice eventually ensures that our results are representative and based on a sufficient number of vulnerabilities, as all metrics in our experiments are computed separately for each \ac{CWE} type.
	
	\textbf{Results.} Using the methodology provided above, we retrieved 794 vulnerable files across three \ac{CWE} types. For additional statistics, see Table \ref{tab:data_stats}. These types, ordered by prevalence of single-file patches in the \ac{CVE} catalog, are:
	\begin{itemize}
		\item CWE-79: \ac{XSS}, the second most dangerous\footnote{The level of danger presented by a particular \ac{CWE} is determined by multiplying the severity score by the frequency score.} \ac{CWE} type in 2022 and the category with the highest number of single-file patches.
		\item CWE-89: SQL Injection, the third most dangerous \ac{CWE} type.
		\item CWE-22: Path Traversal, ranked eighth, but the category with the third highest number of single-file patches; identified as a rarer or 'tail' weakness by \cite{DBLP:conf/kbse/ZhouKXLHL23}.
	\end{itemize}
	
	\begin{table}
		\centering
		% \resizebox{\linewidth}{!}{
			\begin{tabular}{crrrrr|rr} 
				\hline
				\textbf{CWE-ID} & \multicolumn{1}{c}{\begin{tabular}[c]{@{}c@{}}\textbf{No.}\\\textbf{Files}\end{tabular}} & \multicolumn{1}{c}{\begin{tabular}[c]{@{}c@{}}\textbf{Func. }\\\textbf{per File}\end{tabular}} & \multicolumn{1}{c}{\begin{tabular}[c]{@{}c@{}}\textbf{Func. }\\\textbf{Size}\end{tabular}} & \multicolumn{1}{c}{\begin{tabular}[c]{@{}c@{}}\textbf{File }\\\textbf{Size}\end{tabular}} & \multicolumn{1}{c|}{\begin{tabular}[c]{@{}c@{}}\textbf{File Size}\\\textbf{Quartiles}\end{tabular}} & \multicolumn{1}{c}{\begin{tabular}[c]{@{}c@{}}\textbf{Danger }\\\textbf{Score}\end{tabular}} & \multicolumn{1}{c}{\begin{tabular}[c]{@{}c@{}}\textbf{NVD }\\\textbf{Count}\end{tabular}}  \\ 
				\hline
				CWE-22          & 105                                                                                      & 8                                                                                              & 1,073                                                                                       & 7,534                                                                                     & (3,715-14,839)                                                                                      & 14.11                                                                                        & 1,010                                                                                      \\
				CWE-89          & 146                                                                                      & 8                                                                                              & 948                                                                                       & 6,255                                                                                     & (3,213-19,007)                                                                                      & 22.11                                                                                        & 1,263                                                                                      \\
				CWE-79          & 543                                                                                      & 9                                                                                              & 1,005                                                                                       & 8,654                                                                                     & (3,819-17,331)                                                                                      & 45.97                                                                                        & 4,740                                                                                      \\
				\hline
			\end{tabular}
			% }
		\caption{Dataset Statistics: file counts, median file and function sizes (in chars), global statistics for 2022 (frequency and danger of \ac{CWE} in the National Vulnerabilities DB).} \label{tab:data_stats}
	\end{table}
	
	All these \ac{CWE} types fall under the group of \doublequote{error prone processing of data originating from untrusted sources that often results in an initial entry point for an attacker to compromise an IT system}.
	Eventually, our final dataset comprises 794 vulnerable files and the same 794 files with the vulnerability patched (as provided in the CVE catalog), for a total of 1,588 unique files.
	
	\section{RQ1: LLMs vs. In-File Vulnerability Localization} \label{sec:rq1} \label{sec:rq1:benchmark}
	
	The first research question aims to benchmark popular chat-based LLMs on in-file vulnerability localization tasks.
	
	\textbf{Methodology.} For our analysis, we selected three cutting-edge open-source models (Mixtral 8x7b, Mixtral 8x22b, and Llama 3 70b) and three well-known commercial \acp{LLM} from OpenAI: ChatGPT 3.5 \edit{(version 0125)}, ChatGPT 4 \edit{(version 2024-04-09)}, and ChatGPT 4o \edit{(version 2024-05-13)}. 
	The Mistral models have gained significant popularity as the first state-of-the-art open-source \acp{LLM} licensed under Apache 2.0. Llama 3, on the other hand, is the flagship open-source \ac{LLM} from Meta, widely recognized in the AI community and arguably one of the most effective LLMs for vulnerability detection in the open-source landscape \cite{liu2024vuldetectbench,bhusal2024secure}, as well as the most mainstream. Finally, the ChatGPT models, with over 180 million users, are the most widely used commercial LLMs on the market and are the first choice for many developers, as they consistently outperform other general-purpose chat-based LLMs in many (if not most) vulnerability detection benchmarks \cite{liu2024vuldetectbench,steenhoek2024comprehensive,bhusal2024secure,ullah2024llms}.
	For more details about the selected \acp{LLM}, including their size and context window, please refer to Table \ref{table:model_specifications}.
	
	\begin{table}
		\centering
		% \resizebox{\linewidth}{!}{
			\begin{tabular}{lcrcrr} 
				\hline
				\multicolumn{1}{l}{\textbf{Model}} & \begin{tabular}[c]{@{}c@{}}\textbf{Know. }\\\textbf{Cut-off}\end{tabular} & \multicolumn{1}{c}{\textbf{Size}} & \begin{tabular}[c]{@{}c@{}}\textbf{Open}\\\textbf{Source}\end{tabular} & \multicolumn{1}{c}{\begin{tabular}[c]{@{}c@{}}\textbf{Max. }\\\textbf{Tokens}\end{tabular}} & \multicolumn{1}{c}{\begin{tabular}[c]{@{}c@{}}\textbf{Max.}\\\textbf{Characters}\end{tabular}}  \\ 
				\hline
				mixtral-8x7b                       & 2023-12                                                                   & 45B                               & Yes                                                                    & 32k                                                                                         & \textcolor[rgb]{0.208,0.216,0.251}{\textasciitilde{}120k}                                       \\
				mixtral-8x22b                      & 2023-12                                                                   & 141B                              & Yes                                                                    & 64k                                                                                         & \textcolor[rgb]{0.208,0.216,0.251}{\textasciitilde{}250k}                                       \\
				llama-3-70b                        & 2023-12                                                                   & 70B                               & Yes                                                                    & 8k                                                                                          & \textcolor[rgb]{0.208,0.216,0.251}{\textasciitilde{}30k}                                        \\
				gpt-3.5-turbo                      & 2021-09                                                                   & -                                 & No                                                                     & 16k                                                                                         & \textcolor[rgb]{0.208,0.216,0.251}{\textasciitilde{}60k}                                        \\
				gpt-4-turbo                        & 2023-12                                                                   & -                                 & No                                                                     & 128k                                                                                        & \textcolor[rgb]{0.208,0.216,0.251}{\textasciitilde{}500k}                                       \\
				gpt-4o                             & 2023-10                                                                   & -                                 & No                                                                     & 128k                                                                                        & \textcolor[rgb]{0.208,0.216,0.251}{\textasciitilde{}500k}                                       \\
				\hline
			\end{tabular}
			% }
		\caption{Selected \acp{LLM}. This table provides several statistics for each model.}
		\label{table:model_specifications}
	\end{table}
	
	Throughout our experiments, we used these models in their default configurations without any fine-tuning. This approach was chosen to assess the general applicability and effectiveness of these models as they are, i.e., without customization, similar to how a typical software developer might use them. Indeed, our goal is to investigate whether commonly used LLMs can effectively analyze large file-sized inputs, to provide timely insights for software developers and engineers.
	Additionally, by avoiding model-specific fine-tuning, we can also understand the baseline capabilities of these \acp{LLM} in vulnerability detection without the confounding effects of specialized training, which may not be fully replicable or practical in many real-world scenarios.
	
	The only default configuration hyperparameter we changed for all the \acp{LLM} was their \textit{temperature}. The temperature of an \ac{LLM} is a hyperparameter that controls the randomness of its output. It can have any positive real value, with lower values making the output more focused and deterministic, and higher values making it more diverse and creative.
	We set the temperature to 0, the lowest possible value. Setting an \ac{LLM}'s temperature to 0 for vulnerability detection ensures consistent, deterministic, and repeatable responses, which is critical for accurately identifying and analyzing security flaws with little unwanted variability in the results.
	
	With each \ac{LLM}, we used the same prompt, engineered to provide an example of the expected output. This prompting strategy is called \emph{in-context learning} \cite{marvin2023prompt,zhou2024large}. Differently from \cite{DBLP:journals/corr/abs-2401-15468}, we always used the same example across all tested vulnerability types, specifically a CWE-79 bug involving a 'user\_input' JavaScript variable being concatenated into HTML content without proper sanitation.
	
	This is the prompt we used:
	\begin{quoting}
		Analyze the file content below and tell me if there's any line that may contain a bug of type CWE-\{bug\_type\_id\} (\{bug\_type\_label\}). Your output must adhere to the following structure.\\\\
		Expected Output Structure:\\
		SE: very Short Explanation of why the line may contain a bug of given type (e.g., The 'user\_input' is directly concatenated into HTML content without sanitation).\\
		BL: the Bugged Line, if any is found, else none (e.g., `response = "<html><body><h1>Welcome, " + user\_input + "!</h1></body></html>"`).\\
		BUG FOUND: YES if a bug is found, else NO.\\\\
		Example output:\\
		SE: The 'user\_input' is directly concatenated into HTML content without sanitation.\\
		BL: `response = "<html><body><h1>Welcome, " + user\_input + "!</h1></body></html>"`\\
		BUG FOUND: YES\\\\
		File Content:\\
		\{file\_content\}
	\end{quoting}
	
	The required output structure comprises a yes-or-no answer regarding whether the file contains a vulnerability, followed by a brief explanation detailing the rationale behind the potential bug presence in the specified line. For instance, the explanation might note that the 'user\_input' variable is directly concatenated into HTML content without proper sanitation. Additionally, the \ac{LLM} is instructed to also copy and paste the identified problematic line in the output.
	
	The designed prompt explicitly requires the models to analyze the contents of the file and determine if any line might contain a particular bug of the \ac{CWE} type initially associated (in the \ac{CVE} catalog) with the file. For example, for the \doublequote{SshResourceForm.php} file of CVE-2022-24715 shown in Figure \ref{fig:cve_fix}, either before or after the patch, the \acp{LLM} are asked only if any CWE-22 type of vulnerability is present. They are not asked about CWE-89 or CWE-79.

    Following prior work, we measure performance using top-1 accuracy, a standard metric for multi-line vulnerabilities [2,3,4]. A true positive occurs when the model correctly answers "yes" and identifies at least one vulnerable line. "yes" answers without matching lines are false positives, while incorrect "no" answers are false negatives.
    
	\edit{To evaluate whether an \ac{LLM} correctly identified a vulnerability, we followed prior work and measured performance using top-1 accuracy---a standard metric for multi-line vulnerabilities \cite{fu2022linevul,peng2023ptlvd,nguyen2024context}---by examining the yes-or-no answer provided by the model.}
    If the model incorrectly answers \singlequote{yes}, we classify the output as a false positive. Conversely, if the model incorrectly answers \singlequote{no}, we classify the output as a false negative. A correct \singlequote{no} answer is marked as a true negative. For all other cases, differently from \cite{DBLP:journals/corr/abs-2401-15468}, we compare the line content reported by the model to the patch, ensuring normalization of newlines, white-spaces, and tabs. \edit{If the model identifies at least one line that exists in the patch diff (i.e., a line that was changed as part of the bug fix), it means the model correctly detected the buggy line. This is classified as a true positive. Else, if the model fails to detect any line that appears in the patch diff, the output is classified as a false negative.}
	
	\textbf{Results.} Table \ref{table:performance_summary} summarizes the accuracy, F1 scores, precision, and recall across all the considered \ac{CWE} and \acp{LLM}. These scores underline the following finding:
	
	\begin{tcolorbox}[boxsep=1mm, top=1mm, bottom=1mm]
		\textbf{
			Finding 1: Off-the-shelf \acp{LLM} demonstrate low and uneven accuracy across vulnerability types.
		}
	\end{tcolorbox}
	
	\begin{table}[htb]
		\centering
		% \resizebox{\linewidth}{!}{
			\begin{tabular}{llrrrr|r} 
				\hline
				\textbf{CWE-ID}         & \textbf{Model} & \multicolumn{1}{l}{\textbf{F1}}   & \multicolumn{1}{l}{\begin{tabular}[c]{@{}l@{}}\textbf{Accu}\\\textbf{racy}\end{tabular}} & \multicolumn{1}{l}{\begin{tabular}[c]{@{}l@{}}\textbf{Preci}\\\textbf{sion}\end{tabular}} & \multicolumn{1}{l|}{\begin{tabular}[c]{@{}l@{}}\textbf{Rec}\\\textbf{all}\end{tabular}} & \textbf{Files}  \\ 
				\hline
				\multirow{6}{*}{CWE-22} & mixtral-8x7b   & .305          & .195                                                                                     & .268                                                                                      & .352                                                                                    & 210             \\
				& mixtral-8x22b  & .402          & .290                                                                                     & .347                                                                                      & .476                                                                                    & 210             \\
				& llama-3-70b    & .360          & .235                                                                                     & .309                                                                                      & .430                                                                                    & 200             \\
				& gpt-3.5-turbo  & .339          & .212                                                                                     & .292                                                                                      & .404                                                                                    & 208             \\
				& gpt-4-turbo    & \textbf{.470} & \textbf{.324}                                                                            & \textbf{.387}                                                                             & \textbf{.600}                                                                           & 210             \\
				& gpt-4o         & .448          & .295                                                                                     & .368                                                                                      & .571                                                                                    & 210             \\ 
				\hline
				\multirow{6}{*}{CWE-89} & mixtral-8x7b   & .281          & .181                                                                                     & .250                                                                                      & .322                                                                                    & 287             \\
				& mixtral-8x22b  & .362          & .240                                                                                     & .312                                                                                      & .432                                                                                    & 292             \\
				& llama-3-70b    & .400          & \textbf{.276}                                                                            & \textbf{.343}                                                                             & .481                                                                                    & 257             \\
				& gpt-3.5-turbo  & .386          & .255                                                                                     & .328                                                                                      & .469                                                                                    & 286             \\
				& gpt-4-turbo    & .383          & .250                                                                                     & .325                                                                                      & .466                                                                                    & 292             \\
				& gpt-4o         & \textbf{.410} & .260                                                                                     & .341                                                                                      & \textbf{.514}                                                                           & 292             \\ 
				\hline
				\multirow{6}{*}{CWE-79} & mixtral-8x7b   & .173          & .126                                                                                     & .164                                                                                      & .183                                                                                    & 1084            \\
				& mixtral-8x22b  & .220          & .161                                                                                     & .205                                                                                      & .236                                                                                    & 1084            \\
				& llama-3-70b    & .226          & .200                                                                                     & .219                                                                                      & .233                                                                                    & 985             \\
				& gpt-3.5-turbo  & .197          & .217                                                                                     & .202                                                                                      & .192                                                                                    & 1061            \\
				& gpt-4-turbo    & .246          & \textbf{.253}                                                                            & \textbf{.248}                                                                             & .243                                                                                    & 1086            \\
				& gpt-4o         & \textbf{.275} & .184                                                                                     & .247                                                                                      & \textbf{.309}                                                                           & 1086            \\
				\hline
			\end{tabular}
			% }
		\caption{Performance summary by CWE and model. For each CWE, the best scores column-wise are in bold.}
		\label{table:performance_summary}
	\end{table}
	
	The data shows that commercial models generally exhibit superior performance metrics in comparison to their open-source counterparts. Specifically, the models ChatGPT-4 (turbo) and ChatGPT-4o consistently achieve the highest scores across all metrics for CWE-22 and CWE-79. For CWE-89, however, the open-source model Llama 3 surfaces as a strong competitor, in terms of accuracy and precision, surpassing the commercial models except ChatGPT-4o in recall and F1.
	
	Focusing on the open-source models, Llama 3 outperforms the Mixtral series in all evaluated categories and across each \ac{CWE} type, asserting its efficacy among non-commercial options. While the Mixtral 8x22b demonstrates a noticeable improvement over Mixtral 8x7b, particularly in F1 scores and precision, it still falls short of the performance standards set by Llama 3. %This analysis not only highlights the competitive edge of certain commercial models but also underscores the significant potential and efficiency of advanced open-source models in handling complex computational evaluations.
	
	Despite variations in performance across different models and CWE categories, the results suggest that none of the models achieve particularly high metrics, indicating a room for significant improvement. For example, the highest accuracy recorded across all models and categories is only $.324$, a value for the commercial model ChatGPT-4 (turbo) in CWE-22, which is relatively modest in fields where high accuracy and reliability are crucial. Similarly, the best F1 score observed is $.470$, also by ChatGPT-4 (turbo) for CWE-22, suggesting that even the best model's ability to balance precision and recall is limited.
	The open-source models, particularly Llama 3, show promising results but still do not reach a level of performance that could be considered highly effective, with their best accuracy peaking at $.276$ in CWE-89.

    \edit{Since LLMs are known to perform differently across programming languages \cite{buscemi2023comparative,hou2024comparing}, we also provide an analysis based on the programming language distribution encountered in our dataset. The dataset comprises PHP (1020 instances), JavaScript (214 instances), Python (112 instances), and Java (96 instances) files, with smaller contributions from Go, Ruby, and others. A statistical analysis of LLM performance across the top four languages using the Kruskal–Wallis test revealed no significant differences in correctness. In particular, the following results were obtained: Mixtral-8x7b yielded \(H = 1.661\) with \(p = .436\); mixtral-8x22b (\(H = 1.258; p = .533\)); llama-3-70b (\(H = 3.122; p = .210\)); gpt-3.5-turbo (\(H = 1.140; p = .566\)); gpt-4-turbo (\(H = .936; p = .626\)).}
	
	\section{RQ2: The Impact of Vulnerability Position and File Size} \label{sec:rq2}
	
	Our study examines how the location of vulnerabilities within a file and the file size influence the performance of state-of-the-art \acp{LLM} in detecting the three \ac{CWE} types identified in Section \ref{sec:rq1}.
	This section details the two experiments conducted to address RQ2 as outlined in Section \ref{sec:study_design}.
	
	\subsection{File Size and Position of Bug vs. Probability of Detection on Real-World Files} \label{sec:rq2:exp1}
	
	The first experiment examines the correlation between input size, the position of the vulnerability, and the ability of the \ac{LLM} to detect the vulnerability within the 794 vulnerable files in our dataset.
	
	\textbf{Methodology.} Similarly to the previous experiments, also in this case we fed hundreds of vulnerable files to the selected LLMs, asking them to detect and locate the vulnerabilities based on their CWE type. We then studied how well those LLMs did the job, performing an analysis to see whether there is a correlation between their performance and the location of the vulnerabilities as well as the size of the file.
	
\begin{figure*}[htb]
		\centering
		\begin{subfigure}[h]{0.18\linewidth}
			\includegraphics[width=\linewidth]{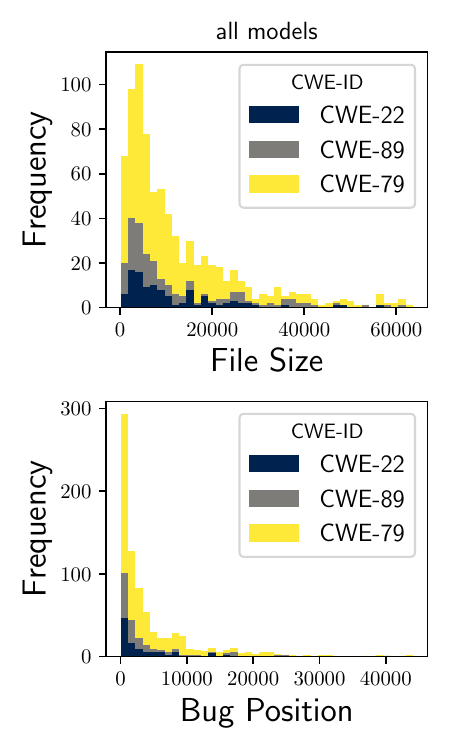}
			%			\caption{Data Distributions}
		\end{subfigure}
		\hfill
		\begin{subfigure}[h]{0.81\linewidth}
			\includegraphics[width=\linewidth]{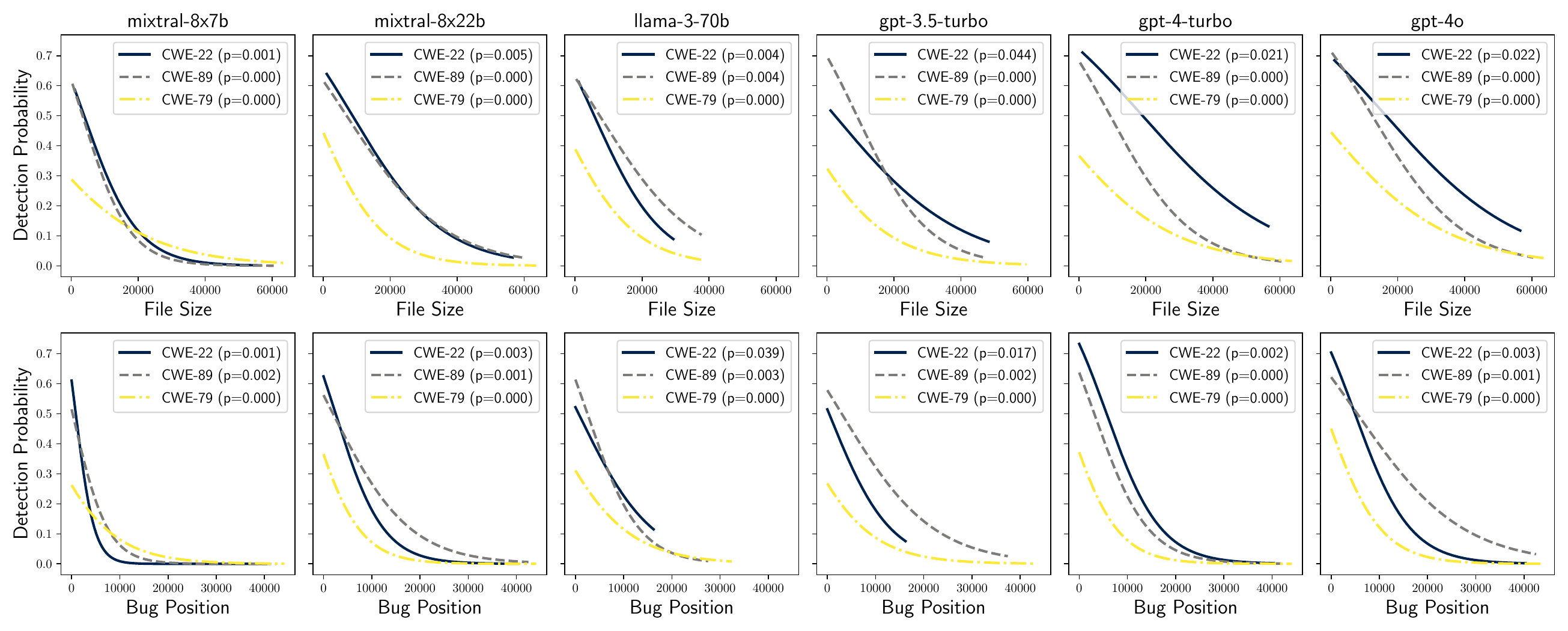}
			%			\caption{Logistic Regressions}
		\end{subfigure}%
		\caption{Real-World Files: Logistic regressions across \acp{LLM} and \ac{CWE} types; data distributions on the left.}
		\label{fig:logistic_regression_all_models}
	\end{figure*}

	Since the LLM's output can only be correct or incorrect, for the correlation analysis we employed logistic regressions, segregating the results by \ac{CWE} type.
    \edit{Specifically, we used \textit{simple logistic regression} models \cite{peng2002introduction} to analyze vulnerability detection as the dependent variable, with either bug position or file size as independent predictors. We chose logistic regression over ANOVA or the Mann-Whitney U test due to the dichotomous nature of the dependent variable and non-normal distribution of predictors (Figure \ref{fig:logistic_regression_all_models}, left).}

    \edit{The test statistic for assessing the significance of each logistic regression coefficient is a z-test, as is typically the case \cite{demaris2013logistic}. In our replication package \cite{anonymous2024infile}, we follow APA guidelines to report the regression coefficients, 95\% confidence intervals, effect sizes (i.e., odds ratios), and p-values for each model term (intercept and predictors). Additionally, we also include the results of a \textit{multiple logistic regression} (i.e., combining both predictors), which remain consistent with those obtained from \textit{simple logistic regression}.}
	
	We measured file size by the total number of characters and determined the position of the vulnerability by counting the characters preceding the vulnerability (as identified through the patch). Although \acp{LLM} process inputs by breaking them into tokens, each \ac{LLM} uses a distinct tokenization strategy. Therefore, to maintain comparability across distinct \acp{LLM}, we opted to measure using characters rather than tokens. \edit{Indeed, measuring effects by character count implies effects in token and line counts, as both are composed of characters.}
	
	\textbf{Results.} The results of the logistic regressions, presented in Figure \ref{fig:logistic_regression_all_models}, reveal that vulnerabilities positioned deeper or within larger files are significantly less likely to be identified by \acp{LLM}. These trends persist across all examined \ac{CWE} types, with p-values significantly below $.05$ in all instances.
	
	% Side by side subfigures 
	
	A notable pattern emerges when combining these logistic regressions with the \ac{CWE} vulnerability frequency in the \ac{CVE} catalog, as shown in Table \ref{tab:data_stats}. 
	The data indicate that the most frequently occurring \ac{CWE} type in our dataset (CWE-79) experiences the deepest decline in vulnerability localization probability. Additionally, in the commercial models (the GPTs), we observe that the negative impact of file size on detection probability is proportional to the frequency of the \ac{CWE} type. Specifically, the line representing CWE-22 (the least frequent type in our dataset) is generally above that of CWE-89, which is above CWE-79.
	%A similar but more pronounced trend is observed in open-source models, particularly with the less frequent CWE-22 type showing a steeper drop compared to the second less frequent CWE-89. 
	
	These trends mirror the observed accuracy scores (see Table \ref{table:performance_summary}). 
	In other words, the trends observed by \cite{DBLP:conf/kbse/ZhouKXLHL23} and mentioned by \cite{DBLP:journals/corr/abs-2401-15468} where frequent vulnerabilities are more easily detected/addressed do not hold true in our specific experiments and are actually inverted.
	%We also found that all the considered models are actually highly sensitive to bug position. This also suggests a potential strategy for improving \ac{LLM} performance: limiting the input size could enhance accuracy.
	
	%We found out with statistical significance that the more frequent vulnerabilities are those less likely to be identified with the probability of a false negative significantly correlating with the file size and the bug position. In other words, the trends observed by \cite{DBLP:conf/kbse/ZhouKXLHL23} and mentioned by \cite{DBLP:journals/corr/abs-2401-15468} do not hold true and are actually inverted.

	\subsection{Code-in-the-Haystack Experiment} \label{sec:rq2:exp2}
	
	Given that the probability of randomly identifying a vulnerable line decreases with increasing bug location and file size and that (as shown in Figure \ref{fig:logistic_regression_all_models}) the bug location and file size follow a tailed distribution, we conducted a follow-up experiment. This second experiment aims to determine whether the previously shown results are due to the inability of the \acp{LLM} to effectively process contextual information or to the tailed distribution of vulnerability features.
	
	\textbf{Methodology.} In the second experiment, called the \singlequote{code-in-the-haystack} experiment, we focused on a limited number of different vulnerability instances, specifically five per \ac{CWE} type. For each instance, we relocated the vulnerable lines of code within the same file, varying the file sizes to create over 500 different file combinations per \ac{CWE} type, to further verify whether the LLMs' detection capabilities correlate with file size and bug position. \edit{Similarly to the other RQ2 experiment, we then used the same \textit{simple logistic regressions} to analyze vulnerability detection.} This experiment was named after the \singlequote{needle-in-the-haystack} experiments discussed in Section \ref{sec:background}.
	%	In particular, the code-in-the-haystack experiment is designed to isolate the effects of file size and vulnerability position on the \ac{LLM}'s vulnerability detection capabilities. 

    In other words, our code-in-the-haystack experiment involves selecting a set of vulnerabilities for each considered \ac{CWE} type and systematically altering the position of the vulnerability's core line within the file so that the distribution of bug positions is uniform. 
    \edit{While in the first RQ2 experiment we examined potential effects at scale, in this second RQ2 experiments we focus on causal analysis, isolating bug position and file size. By design, this experiment holds all potential confounding factors constant, thus isolating the bug position’s impact. In other words, controlling bug position while keeping files unchanged ensured other properties remained fixed.}
	Specifically, in our approach to constructing the experiment, we adopted a methodology consisting of three primary steps. 
	
	In the first step, for each file, we identify and isolate the main line where the vulnerability occurs. If this vulnerable line resides within a function containing fewer than 500 characters, the entire function is marked as the block of vulnerable code to be moved. Conversely, if the function exceeds 500 characters, a segment encompassing a few lines before and after the vulnerable line is extracted. This segment is then refactored into a new function of approximately 500 characters, which is the block of vulnerable code to be moved.
	
	Let $S$ be the size of the file, the second step involves creating $\frac{S}{500}$ new files, positioning the buggy function at different intervals: at the beginning for the first file, at the 500th character for the second, and so on, increasing by 500 characters for each subsequent file. The remaining space in each file is filled with padding characters, using either the rest of the file or contents from files in the same repository.
	
	%	\fixme{Add an equivalent of the Figure 2 (Inputs construction) from \cite{DBLP:journals/corr/abs-2402-14848} to show how we constructed the inputs for the code in the haystack experiment.}
	
	For the final step, we employ the 0/1 knapsack algorithm \cite{pisinger1997minimal} to construct $\frac{S}{500}$ new files, each of size $S$, wherein the buggy function is positioned at different locations. The content of these new files is divided into two segments: one before the buggy function and one after it. Each padding segment is treated as a separate 'knapsack' with capacities determined by the function's position. For example, if the function is placed at position $n$, the first segment's capacity is $500 \times (n-1)$, and the second is $S-500 \times n$. 
	
	To implement this, the original file content is divided into functions or independent logic units, which are then distributed into these knapsacks as padding. Each logic unit is measured in characters. To address cases where the padding does not perfectly match the knapsack's capacity constraints, we use a relaxed version of the 0/1 knapsack algorithm. Instead of requiring an exact match to the capacity, we aim for the best fit possible that is closest to the target capacity, whether it is slightly under or over. We define a tolerance value that allows the total size of selected functions to exceed the capacity by a small amount, i.e., approximately 200 characters.
	
	The aforementioned procedure, however, presents several challenges. A primary challenge is the need to refactor the vulnerable file to allow the free movement of a block code throughout the file. This often requires a deep understanding of both the vulnerability and the underlying repository. Another significant challenge is finding suitable padding content when the target file size exceeds the actual file size.
	Using a \ac{LLM} to address these challenges was excluded, as it could introduce significant bias into the experiment. For instance, the evaluated \ac{LLM} might be already familiar with the code generated for the padding. 
	
	Given the experimental setup requires submitting hundreds of inputs to the \ac{LLM} for each vulnerable file, considering all 794 vulnerable files would substantially increase both the costs and the duration of the experiment. Therefore, we needed to identify the smallest number of files necessary to achieve significant results.
	
	To determine this, we conducted an \textit{a priori} logistic regression power analysis \cite{erdfelder1996gpower}. This analysis considered a one-tailed test, focusing on a strong negative association between bug localization rate and bug position. It assumed a uniform distribution of the predictor, meaning each bug position in a file is equally likely.
	Additionally, the analysis aimed for a power $\beta \geq .8$ with a significance level (\(\alpha\)) of $.05$. Under the null hypothesis, the probability of finding a bug given its position \(Pr(Y=1|X=1)\) was set to $.5$. The analysis also accounted for an \(R^2\) value of $.2$ for other predictors, indicating a small to moderate impact of file content. Given the results of the previous experiments, a medium to large effect size was assumed, corresponding to an odds ratio greater than 3.47 \cite{chen2010big}.
	
	The \textit{a priori} power analysis suggested a minimum sample size of 273 per logistic regression to achieve the desired statistical power. Since each logistic regression considers \(\frac{S}{500}\) files, and the maximum \(S\) we can consider is less than 30,000 characters due to the context window limitation of Llama 3 (the \ac{LLM} with the smallest context window; see Table \ref{table:model_specifications}), we need at least \(\frac{273 \cdot 500}{\sim 30,000} \approx 5\) different vulnerability instances per \ac{CWE} type.
	
	As a result, we selected only 5 different instances of vulnerability per \ac{CWE} type, limiting the scope to \(5 \times 3 = 15\) vulnerable files instead of using the entire dataset of 794.
	
	For the code-in-the-haystack experiment, we considered \(S=\{4,000, 8,000, 16,000, 25,000\}\) characters, resulting in 530 different combinations of files and positions constructed for each \ac{CWE} type and logistic regression, amounting to 28,830,000 characters in total. We chose 25,000 characters as the largest \(S\) since it is slightly lower than the maximum input size of Llama 3. This allowed us to run the same experiments across all six \acp{LLM}.
	
	Each file was processed individually by each \ac{LLM} using the same prompt and matching strategy described in Section \ref{sec:rq1}.
	Given that all the considered \acp{LLM} are non-deterministic and that we have (only) 5 distinct vulnerabilities for each position, file size, and \ac{CWE} type, we conducted the experiment five times with each \ac{LLM} to minimize variability in the results.
	
	To evaluate the results, we assigned a score of +1 when a file's vulnerability was correctly located and -1 when it was incorrectly located. For each run, file size $S$, bug position $n$, \ac{LLM}, and \ac{CWE} type, we averaged the scores across the (5) different instances of vulnerabilities, resulting in an average number within the range of $[-1, 1]$. This score represents the effectiveness of the \ac{LLM} at locating vulnerabilities of the given CWE type. We then averaged these results across multiple runs to achieve more reliable and consistent outcomes.
	
\begin{figure*}[htb]
		\centering
		\includegraphics[width=\linewidth]{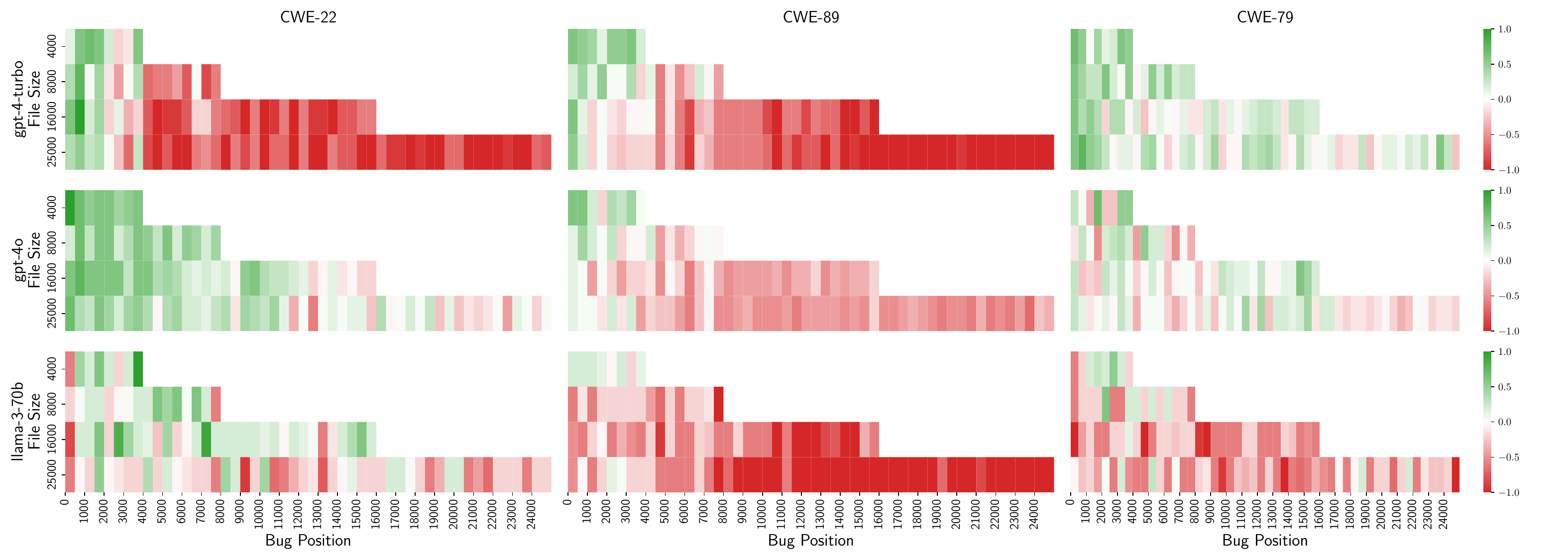}
		%		\caption{Comparative logistic regression analysis across different \acp{LLM}, showing the influence of file size and bug position on the probability of detecting vulnerabilities.}
		\caption{Code-in-the-Haystack: Heatmap results for the most accurate \acp{LLM}.}
		\label{fig:heatmap_grid_llama+gpt4_run5}
	\end{figure*}

\begin{figure*}[htb]
		\centering
		\includegraphics[width=\linewidth]{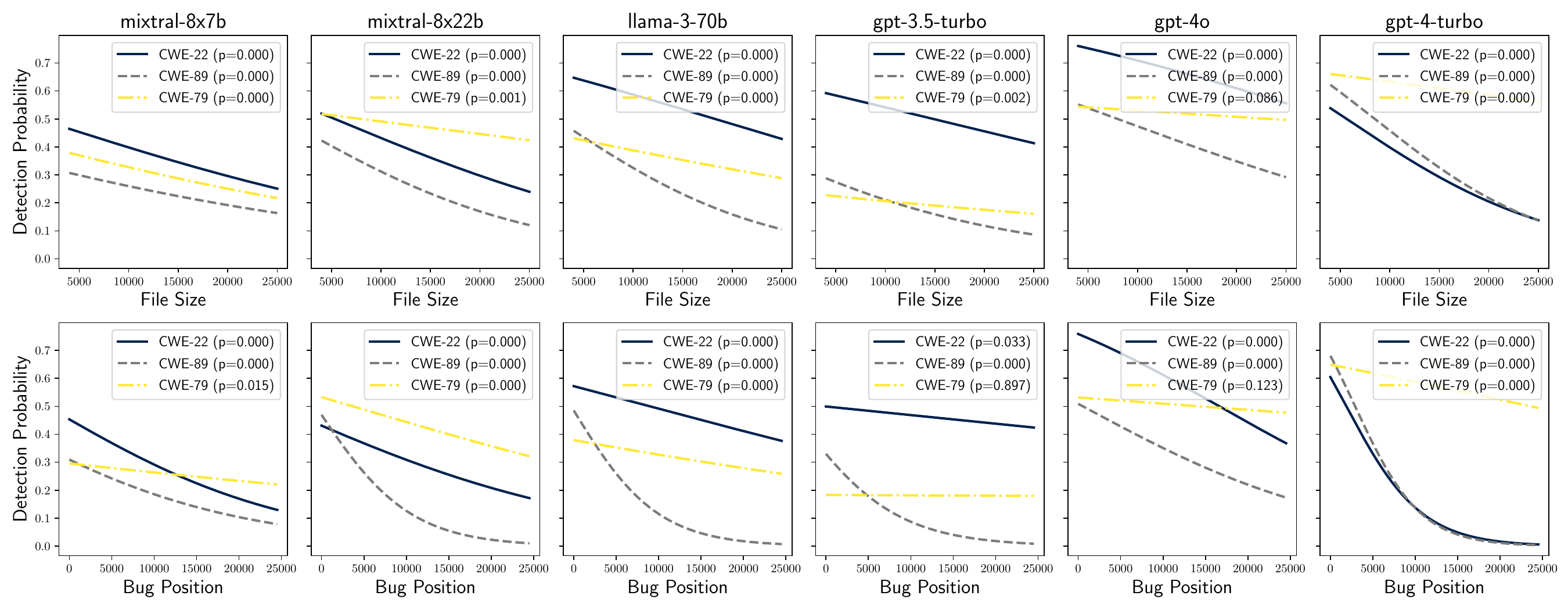}
		%		\caption{Comparative logistic regression analysis across different \acp{LLM}, showing the influence of file size and bug position on the probability of detecting vulnerabilities.}
		\caption{Code-in-the-Haystack: Comparative logistic regression analysis.}
		\label{fig:logistic_regression_all_models_heatmap}
	\end{figure*}

	\textbf{Results.} Figure \ref{fig:heatmap_grid_llama+gpt4_run5} displays the heatmap resulting from the code-in-the-haystack experiment for the top three most accurate models, as indicated in Table \ref{table:performance_summary}: ChatGPT 4, ChatGPT 4o, and Llama 3.
	Meanwhile, Figure \ref{fig:logistic_regression_all_models_heatmap} shows the logistic regressions for all models. These regressions confirm the previous results, indicating a statistically significant negative impact of bug position and file size on vulnerability detection. Interestingly, this trend is more pronounced with bug position (i.e., the logistic regressions are steeper) than with file size, suggesting that position plays a bigger role.
	
	Notably, Figure \ref{fig:heatmap_grid_llama+gpt4_run5} shows that, when dealing with the largest files, the \acp{LLM} predominantly focus on the beginning of the file, disregarding the content towards the end of the context window. This observation indicates that the models do not suffer from a \singlequote{lost-in-the-middle} phenomenon. Instead, they exhibit a \singlequote{lost-in-the-end} problem. The degree to which this trend occurs varies among different models and \ac{CWE} types.
	
	\begin{figure*}[bth]
		\centering
		\includegraphics[width=\linewidth]{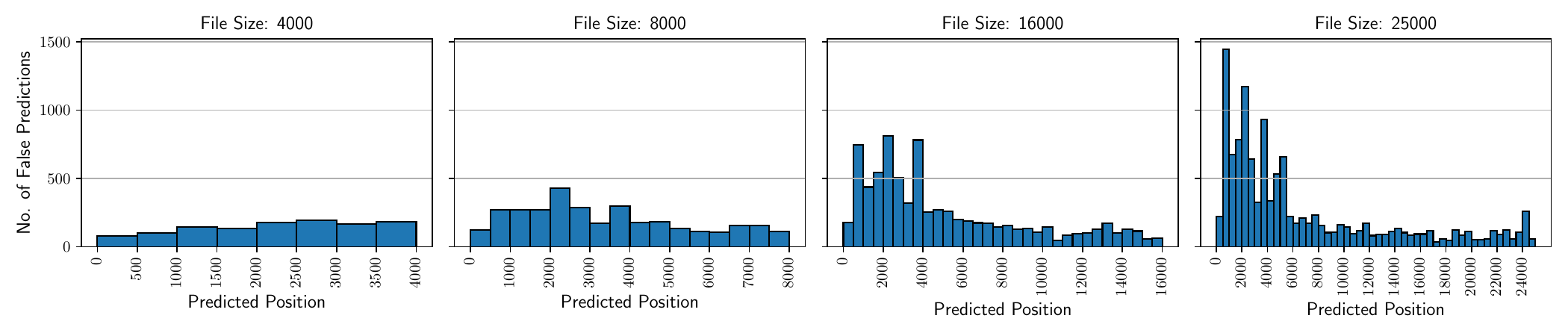}
		%		\caption{Comparative logistic regression analysis across different \acp{LLM}, showing the influence of file size and bug position on the probability of detecting vulnerabilities.}
		\caption{Code-in-the-Haystack: Distribution of incorrectly predicted bug positions.}
		\label{fig:false_negatives_predicted_position_distribution}
	\end{figure*}
	
	To further verify this finding, we conducted an error analysis. As shown in Figure \ref{fig:false_negatives_predicted_position_distribution}, we plotted the distribution of incorrect bug positions predicted by all the \acp{LLM}. The results clearly indicate that the \acp{LLM} tend to predict vulnerabilities at the earliest positions (< 6,000) when they fail to identify a vulnerability, often neglecting positions towards the end of the file.
	
	Our code-in-the-haystack experiment is designed to have a uniform distribution of bug positions for each of the file sizes $S$. If there were no \singlequote{lost-in-the-end} phenomenon, we would expect to see a more-or-less uniform distribution of errors in Figure \ref{fig:false_negatives_predicted_position_distribution}, similar to what we observe for $S$ = \{4,000, 8,000\}. However, for $S$ = \{16,000, 25,000\}, the predicted positions are skewed towards the file's beginning.
	These observations support the following finding:
	\begin{tcolorbox}[boxsep=1mm, top=1mm, bottom=1mm]
		\textbf{Finding 2: \edit{The LLMs we examined struggle to detect security bugs---at least those of types CWE-22, CWE-89, and CWE-79---toward the end of longer files.}}
	\end{tcolorbox}
	
	Figures \ref{fig:heatmap_grid_llama+gpt4_run5} and \ref{fig:false_negatives_predicted_position_distribution} show a threshold beyond which the \acp{LLM} starts struggling with vulnerability detection. For example, ChatGPT 4, 4o, and Llama 3 have difficulty detecting CWE-89 type vulnerabilities when the input exceeds 4,000 characters. The next section explores how to identify this threshold for each \ac{CWE} type and \ac{LLM}.
	
	\section{RQ3: Input Size Identification} \label{sec:rq3}
	
	With RQ3, we aim to identify best practices for practitioners that use popular chat-based LLMs, such as GPT models, which suffer from the \singlequote{lost-in-the-end} issue. Since these LLMs struggle to detect vulnerabilities in files approaching their context window limits, we seek to provide guidance on the right input size for reliable detection.
	%Our objective is to determine the best input size to detect vulnerabilities effectively without experiencing the \singlequote{lost-in-the-end} effect. 
	Specifically, we want to identify the largest input size for a \ac{CWE} type at which the model's detection capabilities have the highest recall. % and do not degrade when the vulnerability is located deeper in the context window.
	We hypothesize that the best input size can be determined by observing changes in the \ac{LLM}'s ability to identify the exact location of a vulnerability when positioned into file chunks of different sizes. To investigate this hypothesis, we use the data from the first experiment of RQ2 (cf. Section \ref{sec:rq2:exp1}), i.e., the 794 vulnerable files.
	
	\textbf{Methodology.} Our experimental procedure involves a naive chunking strategy where the content of a file is divided into blocks of lines totaling up to a maximum of $k$ characters. Each line is preserved in its entirety to ensure at least one chunk contained the critical line for the \ac{LLM} to identify. Although simplistic, this approach demonstrates that even basic chunking could enhance the \ac{LLM}'s performance.% We reserve the exploration of more sophisticated chunking strategies for future studies.
	
	Each chunk was evaluated using the same experimental prompts used in the investigations of RQ1 and RQ2. Due to the increase in negative examples compared to RQ1, direct comparisons of accuracy and F1 scores are not viable. Instead, we focused on comparing the recall metric, representing the percentage of correctly identified vulnerabilities by the \ac{LLM}.
    The chunking strategy was implemented using \( k \) values of \{6,500, 3,000, 1,500, 500\}. \edit{The largest chunk size, 6,500, corresponds to approximately the median file size for CWE-89 (see Table \ref{tab:data_stats}). A chunk size of 3,000 represents the lower quartile of file sizes across all CWE types. The 1,500 size is slightly above the average function size, while 500 is less than the average function size, providing a finer granularity.}
	
	\begin{table}
		\centering
		%\resizebox{\linewidth}{!}{
			\begin{tabular}{clrrrr} 
				\hline
				\multicolumn{1}{l}{\textbf{CWE-ID}} & \textbf{Model} & \multicolumn{1}{l}{\begin{tabular}[c]{@{}l@{}}\textbf{Chunk }\\\textbf{Size}\end{tabular}} & \multicolumn{1}{l}{\begin{tabular}[c]{@{}l@{}}\textbf{Accu}\\\textbf{racy}\end{tabular}} & \multicolumn{1}{l}{\begin{tabular}[c]{@{}l@{}}\textbf{Rec}\\\textbf{all}\end{tabular}} & \multicolumn{1}{l}{\begin{tabular}[c]{@{}l@{}}\textbf{Recall }\\\textbf{Improv.}\end{tabular}}  \\ 
				\hline
				\multirow{6}{*}{CWE-22}             & mixtral-8x7b   & 500                                                                                        & .206                                                                                     & .428                                                                                   & +21.59\%                                                                                        \\
				& mixtral-8x22b  & 3000                                                                                       & .219                                                                                     & .552                                                                                   & +15.97\%                                                                                        \\
				& llama3-70b     & 1500                                                                                       & .197                                                                                     & .495                                                                                   & +15.12\%                                                                                        \\
				& gpt-3.5-turbo  & 1500                                                                                       & .220                                                                                     & .533                                                                                   & \textbf{+31.93\%}                                                                               \\
				& gpt-4-turbo    & 6500                                                                                       & .363                                                                                     & .657                                                                          & +9.50\%                                                                                         \\
				& gpt-4o         & 6500                                                                                       & .320                                                                                     & \textbf{.666}                                                                          & +16.64\%                                                                                        \\ 
				\hline
				\multirow{6}{*}{CWE-89}             & mixtral-8x7b   & 1500                                                                                       & .183                                                                                     & .452                                                                                   & +40.37\%                                                                                        \\
				& mixtral-8x22b  & 500                                                                                        & .122                                                                                     & \textbf{.657}                                                                                   & \textbf{+52.08\%}                                                                               \\
				& llama3-70b     & 1500                                                                                       & .330                                                                                     & .554                                                                                   & +15.18\%                                                                                        \\
				& gpt-3.5-turbo  & 500                                                                                        & .278                                                                                     & .636                                                                          & +35.61\%                                                                                        \\
				& gpt-4-turbo    & 500                                                                                       & .462                                                                                     & .630                                                                                   & +35.19\%                                                                                        \\
				& gpt-4o         & 1500                                                                                       & .238                                                                                     & .650                                                                                   & +26.46\%                                                                                        \\ 
				\hline
				\multirow{6}{*}{CWE-79}             & mixtral-8x7b   & 500                                                                                        & .375                                                                                     & .309                                                                                   & +68.85\%                                                                                        \\
				& mixtral-8x22b  & 500                                                                                        & .147                                                                                     & \textbf{.428}                                                                          & +81.36\%                                                                                        \\
				& llama3-70b     & 1500                                                                                       & .474                                                                                     & .278                                                                                   & +19.31\%                                                                                        \\
				& gpt-3.5-turbo  & 500                                                                                        & .397                                                                                     & .376                                                                                   & \textbf{+95.83\%}                                                                              \\
				& gpt-4-turbo    & 500                                                                                        & .492                                                                                     & .404                                                                                   & +66.26\%                                                                                        \\
				& gpt-4o         & 500                                                                                        & .380                                                                                     & .407                                                                                   & +31.72\%                                                                                        \\
				\hline
			\end{tabular}
			%}
		\caption{Chunk sizes yielding the highest recall for each \ac{LLM} and \ac{CWE} type. Recall improvements are compared to the baseline values from Table \ref{table:performance_summary}. Best values are in bold.}
		\label{tab:chunking_results}
	\end{table}
	
	\textbf{Results.} According to the results detailed in Table \ref{tab:chunking_results}, we have an average recall improvement of over +37\% across all models and \ac{CWE} types due to chunking, although this strategy led to a decrease in precision, indicated by an increase in false positives due to a greater number of chunks with no vulnerability.
    
    For CWE-22 with a 1,500 character chunk, ChatGPT 3.5 showed a significant +31.93\% recall improvement. In contrast, ChatGPT 4 (turbo) with a 6,500 character chunk reached a +9.5\% improvement for the same \ac{CWE}. ChatGPT 3.5 with a 500 character chunk for CWE-79 showed the most substantial improvement, almost doubling the baseline recall (+95.83\%).
	
	Interestingly, the smaller \acp{LLM} demonstrated the most significant improvements in recall and appeared to benefit more from smaller chunk sizes. 
	For example, in CWE-89, Mixtral 8x22b, using a chunk size of 500, achieved a +52.08\% improvement in recall, the highest among all models for this specific \ac{CWE} type.
	Furthermore, non-commercial \acp{LLM} also showed greater performance gains with smaller chunks, narrowing the performance gap with commercial models. In fact, after chunking, open-source models frequently outperformed commercial ones in vulnerability detection.
	
	The data also reveal disparities in how different \ac{CWE} types respond to chunking. For instance, CWE-79 is the most susceptible to \singlequote{lost-in-the-end} issues, benefiting the most from smaller chunk sizes of 500 characters (less than a function!). CWE-89 and CWE-22 followed, showing optimal safe input sizes which are around 1,500 and 3,000-6,500 characters, respectively.
	When we compare these input sizes to the median file and function sizes in Table \ref{tab:data_stats}, we observe notable differences across \ac{CWE} types. Specifically, CWE-22 requires a small-to-medium file size, whereas CWE-89 and CWE-79 require the sizes of a small function and less than half a function, respectively.
	These observations substantiate the following finding:
	\begin{tcolorbox}[boxsep=1mm, top=1mm, bottom=1mm]
		\textbf{Finding 3: Chunking input files into smaller blocks enhances \acp{LLM}'s recall in vulnerability detection, with optimal sizes varying a lot by \ac{CWE} type.}
	\end{tcolorbox}

	\section{Threats to Validity} \label{sec:threats_to_validity}
	
	In evaluating the findings, several extrinsic and intrinsic potential threats to validity need to be considered.
	
	\textbf{Extrinsic Threats.} While we included a variety of popular \acp{LLM}, such as Mixtral and Llama, as well as several versions of ChatGPT, the conclusions drawn are primarily applicable to the contexts and vulnerability types explicitly tested, i.e., CWE-22, CWE-79, and CWE-89. The performance of these models might differ when applied to other \ac{CWE} types or in different software contexts, such as different programming languages. %Our dataset, comprising vulnerabilities with single-file, single-commit patches from GitHub, could benefit from the inclusion of data from other repositories or more complex commit structures to enhance the generalizability of our findings.
	
	OpenAI's continuous updates to their models also pose a challenge; the accuracy and F1 scores presented might change if the experiments are rerun at a different time. Although we aimed for consistency and simplicity in the prompts used, future versions of ChatGPT may perform differently or require new types of prompts. 
	
	Lastly, our analysis using logistic regression to assess the impact of file size and vulnerability position on detection probability assumes a linear relationship, which may not capture more complex patterns adequately. 
	
	\textbf{Intrinsic Threats}.  Generally speaking, it is impossible to mathematically guarantee that a non-trivial piece of code (e.g., a vulnerability fix) is fully bug-free. This impossibility stems from Rice's theorem \cite{rice1953classes}. Nevertheless, we are certain that the post-fix code we used for answering RQ1 resolves at least one instance of vulnerability, and that this vulnerability must be detected by the LLM. Instead, our core findings and contributions, which stem from RQ2 and RQ3, are not affected by this problem.
	
	The algorithmic construction and manual refactoring of our dataset's files for the code-in-the-haystack experiments could have introduced syntactical errors, such as misplaced import statements. However, this is not a major concern as the study focuses on the \singlequote{lost-in-the-end} phenomenon in detecting specific types of security weaknesses rather than on syntactical errors.
	
	The non-deterministic nature of the \ac{AI} models, particularly the non-commercial ones that lack control over the random seed, adds another layer of complexity. To mitigate this, we repeated the code-in-the-haystack experiments five times and averaged the results, incurring costs exceeding 400 USD. However, due to these costs, this approach was not feasible for other experiments, and even five runs may be insufficient. Nonetheless, the consistent observation of the \singlequote{lost-in-the-end} issue across all studied \acp{LLM} and the statistical significance of the logistic regression support the reliability of our findings, reducing the likelihood that the observed trends are due to chance. 
	
	Finally, the range of chunk sizes $k$ we considered is arbitrary, and better chunk sizes might exist. Due to the high costs of running large experiments with \acp{LLM}, we were limited to testing a few $k$ values using a dichotomic search.
	
	\section{Discussion} \label{sec:discussion}
	
	The three research questions outlined in Section \ref{sec:study_design} focused on evaluating the effectiveness of several popular \acp{LLM} in detecting in-file vulnerabilities. Specifically, they examined the influence of file size and the position of the vulnerability, as well as the optimal input size for maximizing detection accuracy.
	
	\textbf{RQ1.} The findings presented in Section \ref{sec:rq1} address RQ1, indicating that the performance of off-the-shelf \acp{LLM} is generally low, with accuracy scores below $.4$, and varies significantly across different \ac{CWE} types. No model consistently outperformed others across all aspects, though commercial models generally show superior performance compared to open-source models. Notably, the majority of the \acp{LLM} were more effective in detecting CWE-22 and CWE-89 compared to CWE-79. This suggests that the complexity and commonality of vulnerabilities influence detection capabilities. Interestingly, our results diverge from previous studies \cite{DBLP:journals/corr/abs-2401-15468,DBLP:conf/kbse/ZhouKXLHL23}, which suggested that more frequent vulnerabilities are easier to detect for these models. Our findings show that the most frequent vulnerabilities can be the hardest to spot for \acp{LLM}.
	
	\textbf{RQ2.} For RQ2, our experiments (cf. Section \ref{sec:rq2}) reveal a clear pattern: as file size increases or the vulnerability is positioned towards the end of the file, the probability of its detection by \acp{LLM} decreases. This finding was consistent across all models and types of vulnerabilities studied, regardless of the \acp{LLM}' maximum context window size. This highlights the challenges \acp{LLM} face with large inputs and their sensitivity to the \singlequote{lost-in-the-end} phenomenon. Additionally, our analysis showed that different \ac{CWE} types are handled differently. For example, CWE-79 had the deepest decline in vulnerability localization probability, suggesting that some vulnerabilities may require more context for accurate identification, but current \acp{LLM} limitations in handling large contexts prevent proper localization.

    \edit{Notably, the distributions of bug positions and file sizes shown in Figure \ref{fig:logistic_regression_all_models} (left) indicate that real-world bugs frequently occur within the first 10,000 characters of a file. The same bar plot also suggests a correlation between bug position and file size, possibly because smaller files inherently constrain bug placement. Indeed, the file size distribution is skewed toward smaller files, with most containing fewer than 20,000 characters. To account for this correlation, we conducted the code-in-the-haystack experiment (Section \ref{sec:rq2:exp2}) using uniformly and independently distributed bug positions and file sizes. The results still indicate a \singlequote{lost-in-the-end} phenomenon.}

    \edit{Importantly, when analyzing the results of the code-in-the-haystack experiment, we observe that the distribution of incorrectly predicted bug positions for files of size $S=$25,000 (see Figure \ref{fig:false_negatives_predicted_position_distribution}, rightmost box) closely resembles the distribution of bug positions in real-world files collected from the CVE catalog. The LLMs incorrectly predict that bug positions occur within the initial 10,000 characters of a file. This suggests that the skewed distribution of real-world bug positions may be why LLMs (likely trained on that data) tend to focus on the beginning of files, using code position as a heuristic for bug localization rather than analyzing the underlying code. This observation aligns with the findings of \citet{nikankin2024arithmetic}, which show that LLMs often rely on a \singlequote{bag of heuristics} to solve problems. If this heuristic reliance is accurate, alternative training strategies may be necessary. For instance, training LLMs with more uniformly distributed bug positions and file sizes or refining the training pipeline could potentially mitigate the \singlequote{lost-in-the-end} phenomenon.}

    \textbf{RQ3.} The results from our chunking experiments addressing RQ3 suggest that smaller input sizes can significantly improve the recall of \acp{LLM} in detecting vulnerabilities (up to +95\%), depending on the \ac{CWE} type. Specifically, smaller chunk sizes of 500 to 1,500 characters were more effective for the most frequent \ac{CWE} types, while larger chunk sizes up to 6,500 characters or more were most effective for the least frequent \ac{CWE} type, CWE-22. This indicates that at least for CWE-22, in-file vulnerability detection is needed.

    Given that the vulnerabilities we are studying rank among the top 10 most dangerous ones (cf. Section \ref{sec:rq1:data}), maximizing recall (i.e., the true positive rate) can be more important than precision or accuracy. Therefore, adopting a chunking strategy like the one described could, with little to no additional cost, increase the vulnerability detection rates of state-of-the-art \acp{LLM}, effectively mitigating their \singlequote{lost-in-the-end} issues.

    \edit{Nonetheless, chunking is a useful but temporary solution for managing large files, often leading to increased alert fatigue---a problem also identified by \citet{hassan2019nodoze}---due to a rise in false positives. When examining this phenomenon, we find that most false positives result from a loss of context caused by chunking. For example, all LLMs incorrectly flag the following as buggy: `query = "SELECT * FROM users WHERE username="+user\_input+""` simply because the `user\_input` sanitization is out of context.}
    
    \edit{A potential solution to this problem is \singlequote{smarter chunking}, where variable definitions and manipulations are included in the chunk. However, this chunking approach could effectively address only input sanitization issues such as SQL injection, path traversal, and XSS. But, given that these are among the most dangerous and common vulnerabilities, employing this specialized strategy may still be worthwhile. Instead, for other types of vulnerabilities, a more effective approach is to address the \singlequote{lost in the end} issue directly within the LLM architecture and training pipeline.}
	
	Although the chunking strategy employed is somewhat naive and encounters similar issues as function-level detection due to limited contextual information, the results (see Table  \ref{tab:chunking_results}) indicate that the most effective chunk sizes can exceed the average function size (shown in Table \ref{tab:data_stats}). This further confirms that function-level detection may be ineffective, particularly for certain vulnerabilities like CWE-22 and, to a lesser extent, CWE-89.
	
	Interestingly, when considering smaller input sizes, the performance gap between commercial and open-source models diminished notably. In some cases, open-source models performed on par with or even better than commercial ones, suggesting that the main advantage of commercial models may lie in a better handling of context windows. This indicates that advancements in handling context windows could significantly improve the existing technology for vulnerability detection. However, it is unclear whether these \acp{LLM} suffer from the \singlequote{lost-in-the-end} problem because they had mostly seen data where vulnerability detection is done predominantly on function-level data.

    \edit{While our findings highlight significant improvements in recall through chunking, they also underscore the need for practitioners to adopt a more nuanced approach when selecting and configuring LLMs for vulnerability detection. Practitioners should consider not only the chunk size but also the nature of the CWE type and the model’s context-handling capabilities. For instance, smaller chunk sizes (500--1,500 characters) yield superior results for CWE types such as CWE-79 and CWE-89, whereas larger chunks (up to 6,500 characters) may be necessary for more context-dependent vulnerabilities like CWE-22. Commercial models like GPT-4-turbo and GPT-4o demonstrate stronger context retention, making them more effective for CWE-22. Instead, open-source models such as Mixtral-8x22b can achieve higher recall performance when smaller context retention is required (e.g., for CWE-79 and CWE-89), outperforming their commercial counterparts, which appear to be more specialized for longer contexts.}
	
	\textbf{Hypotheses.} We hypothesize that the problem is the combination of the context-dependency of certain vulnerabilities and the \singlequote{lost-in-the-end} issue, so that \ac{CWE} types requiring more context generally perform worse. For instance, detecting \ac{XSS} (CWE-79) requires understanding how user inputs are handled, sanitized, and embedded into web pages, which can span various functions, files, or even different programming languages (e.g., JavaScript, HTML). If the model cannot capture the broader context (maybe because it has not seen many large contexts during training), it might miss the vulnerability, as shown by our empirical results. 
	
	Similarly, if the code context is not correctly captured by the model, it may mistakenly identify a non-buggy line as having a vulnerability due to the fix being elsewhere in the file. Hence, contrary to what \cite{DBLP:journals/corr/abs-2401-15468} suggest, it is possible that the harder-to-locate bugs are those with greater variability in the type and quantity of code modifications necessitated by a security patch, rather than the rarer vulnerabilities, i.e., the tail \ac{CWE} types.
	
	Another possibility is that the code used for training these \acp{LLM} frequently contains unidentified bugs, leading the models to replicate them or fail to recognize them as bugs. Indeed, it is possible that the most frequently fixed \ac{CWE} vulnerability types (CWE-79) also more frequently appear unfixed in production or open-source code. If the model cannot handle a large context effectively, it might learn these vulnerabilities as normal non-buggy code, as suggested by the experiments conducted by \cite{DBLP:journals/corr/abs-2403-15600}, which indicate that \acp{LLM} tend to replicate bugs.
	
	\textbf{Future Work.} Future work should verify whether any of the hypotheses mentioned above actually hold. Moreover, given that code, unlike natural language, does not follow a linear, top-down format, future research should explore more advanced architectures for these LLMs. This could potentially include permutation-invariant neural networks, such as graph neural networks, where reordering a function’s position does not affect the model’s ability to detect vulnerabilities.
	
	\section{Conclusion} \label{sec:conclusion}
	
	This paper characterized and highlighted the lost-in-the-end issue in popular chat-based LLMs. Recognizing this issue is important for software developers and engineers who extensively use these specific LLMs for coding and debugging activities such as file-level vulnerability detection.
	
	In this study, we studied the effectiveness and limitations of popular chat-based \acp{LLM} in detecting vulnerabilities within entire source code files, focusing on three of the most dangerous and common \ac{CWE} vulnerability types: CWE-22, CWE-89, and CWE-79. Our findings reveal significant variability in \ac{LLM} performance across these vulnerabilities and highlight a new challenge: the \singlequote{lost-in-the-end} issue, where vulnerabilities located towards the end of files are less likely to be detected.
	
	We showed that the \singlequote{lost-in-the-end} issue can considerably affect the performance of \acp{LLM} on file-level vulnerability detection. This also implies that the same issue could affect \acp{LLM} at similar tasks involving reasoning over large software files such as code review automation, generic bug localization, code summarization.
	
	The implications of our findings are twofold. Firstly, they suggest that further improvements in \acp{LLM} are needed before they can be reliably used for vulnerability detection in software development. This includes enhancements in their ability to handle large inputs and in their sensitivity to the placement of vulnerabilities within files. Secondly, our study highlights the potential of simple yet effective strategies like input chunking to significantly enhance the performance of existing \acp{LLM}, which could be readily applied in current software development practices. %Indeed, our empirical results indicate that smaller chunks can substantially improve model performance in terms of recall. 
	
	Although our analysis (Figure \ref{fig:logistic_regression_all_models}) indicates that the smallest files are the most common, the \singlequote{lost-in-the-end} issue is still relevant because the largest files should not be ignored and \doublequote{the devil is in the tails} \cite{DBLP:conf/kbse/ZhouKXLHL23}. Therefore, to eliminate the most dangerous vulnerabilities, we strongly believe that future research should keep increasing the context window of these \acp{LLM} up to the repository level, but in a way that is more robust to the \singlequote{lost-in-the-end} issue. 

    \section*{Data Availability} 
    All the data and scripts used for this paper are available in our replication package \cite{anonymous2024infile}.
	
	%%
	%% The acknowledgments section is defined using the "acks" environment
	%% (and NOT an unnumbered section). This ensures the proper
	%% identification of the section in the article metadata, and the
	%% consistent spelling of the heading.
	\begin{acks}
		F. Sovrano and A. Bacchelli gratefully acknowledge the support of the Swiss National Science Foundation through the SNF Project 200021\_197227.
	\end{acks}
	
	%%
	%% The next two lines define the bibliography style to be used, and
	%% the bibliography file.
	\bibliographystyle{ACM-Reference-Format}
	\bibliography{bibliography}

	%%
	%% If your work has an appendix, this is the place to put it.
	%\appendix
	%
	%\section{Research Methods}
	%
	
\end{document}